PRE-PRINT with no revision

# Complementary two dimensional carrier profiles of 4H-SiC MOSFETs by Scanning Spreading Resistance Microscopy and Scanning Capacitance Microscopy

Patrick Fiorenza[1,a*], Marco Zignale[1,b], Edoardo. Zanetti[2,c], Mario S. Alessandrino[2,d], Beatrice Carbone[2,e], Alfio Guarnera[2,f], Mario Saggio[2,g], Filippo Giannazzo[1,h] and Fabrizio Roccaforte[1,i]

[1] Consiglio Nazionale delle Ricerche – Istituto per la Microelettronica e Microsistemi (CNR-IMM), Strada VIII 5 95121, Catania, Italy

[2]STMicroelectronics Stradale Primosole 50 95121, Catania, Italy

[a]patrick.fiorenza@imm.cnr.it, [b]marco.zignale@imm.cnr.it, [c]edoardo.zanetti@st.com, [d]santi.alessandrino@st.com, [e]beatrice.carbone@st.com, [f]alfio.guarnera@st.com, [g]mario.saggio@st.com, [h]filippo.giannazzo@imm.cnr.it, [i]fabrizio.roccaforte@imm.cnr.it

**Abstract.** This paper reports the results presented in an invited poster during the International Conference on Silicon Carbide and Related Materials (ICSCRM) 2023 held in Sorrento, Italy. The suitability of scanning probe methods based on atomic force microscopy (AFM) measurements is explored to investigate with high spatial resolution the elementary cell of 4H-SiC power MOSFETs. The two-dimensional (2D) cross-sectional maps demonstrated a high spatial resolution of about 5 nm using the SSRM capabilities. Furthermore, the SCM capabilities enabled visualizing the fluctuations of charge carrier concentration across the different parts of the MOSFETs elementary cell.

## Introduction

New generation of silicon carbide (4H-SiC) metal oxide semiconductor field effect transistors (MOSFETs) are increasing their performance in terms of On-resistance ($R_{ON}$), and maximum operating current also by shrinking of the cell-pitch and optimized thermal activation of the implanted regions [1]. In 4H-SiC MOSFETs, ion implantation is commonly used to introduce dopant species (Phosphorous for n-type and Aluminum for p-type) in confined regions of the semiconductor material, followed by high-temperature annealing for the electrical activation [2,3]. As matter of fact, the designed doping level is chosen by TCAD simulation of the final device structure. However, to understand the real device performance, both the active doping concentration (or the material resistivity) and the real device geometry (e.g. size of the implanted region, junction depths, etc.) must be accurately monitored at the nanoscale.

In the last decade, both scanning capacitance microscopy (SCM) [4] and scanning spreading resistance microscopy (SSRM) [5] demonstrated their capabilities to detect the introduction of electrically active atomic species (i.e. nitrogen and/or phosphorous) by post oxide deposition annealing of the gate insulators. Furthermore, scanning capacitance based methods [6] have been used to get information of the interface state density spatial distribution of MOS systems.

In this context, two dimensional (2D) electrical scanning probe techniques (SPM), such as SCM and SSRM), can give useful information both on the spatial distribution of the active dopants concentration and local resistivity in the region underneath the tip [7]. The SCM and SSRM have been widely employed for quantitative 2D carrier profiling in silicon-based CMOS structures [8]. On the other hand, even though SCM [7,9] and SSRM [10] potentialities have been explored onto wide bandgap semiconductors, dedicated efforts are still needed to assess their 2D profiling capabilities. In particular, the current injection mechanisms at the SPM tip/4H-SiC contact need to be further investigated, to achieve a quantification of SSRM map. Nevertheless, the combined use of SCM and SSRM on the same device structures can provide precious complementary information eventually on the discrepancy between the designed and the real device, or on processing issues.

In this paper, SCM and SSRM analyses have been performed on the channel region of a vertical 4H-SiC power MOSFET, with the aim to determine the n-type and p-type distribution in real device, and to estimate the lateral resolution of these two complementary SPM methods and their capabilities on the detection of doping level variation across the unitary cell.

## Experimental

The discrete MOSFETs devices under investigation were fabricated on 4°-off-axis epitaxial n-type (0001) 4H-SiC layers having a nominal a donor concentration in the range of $10^{16}$ $cm^{-3}$ and a body region formed by an aluminium implantation (nominally about $N_A$~$10^{17}$ $cm^{-3}$), while Phosphorous implantation was used for the source region ($N_D$ ~ $10^{20}$ $cm^{-3}$) [11].

The devices were prepared in a bevelled edge of 5° 44’ enabling the geometrical magnification of the depth profile by a factor 10×. Before SCM analyses, the bevelled sample were subjected to an immersion in $H_2O_2$ at 40 vol. for 20 min to produce the formation of a native oxide on the SiC surface [12]. The 2D imaging analyses were carried out using a DI3100 AFM by Bruker with a Nanoscope V controller equipped with the SCM and SSRM modules. Doped diamond-coated Si tips were employed to ensure electrical stability during the scan of the 4H-SiC surfaces.

## Results and discussion

Fig. 1 shows the TCAD simulated structure of the planar power MOSFET in cross section, where two different cut lines are reported to estimate the free carrier concentration across the channel and the JFET regions. In particular, both the electron (in red) and hole (in blue) distributions are reported in two different regions of the device elementary cell, such as the source-body-drift region (cutline 1) and the JFET-drift region (cutline 2).

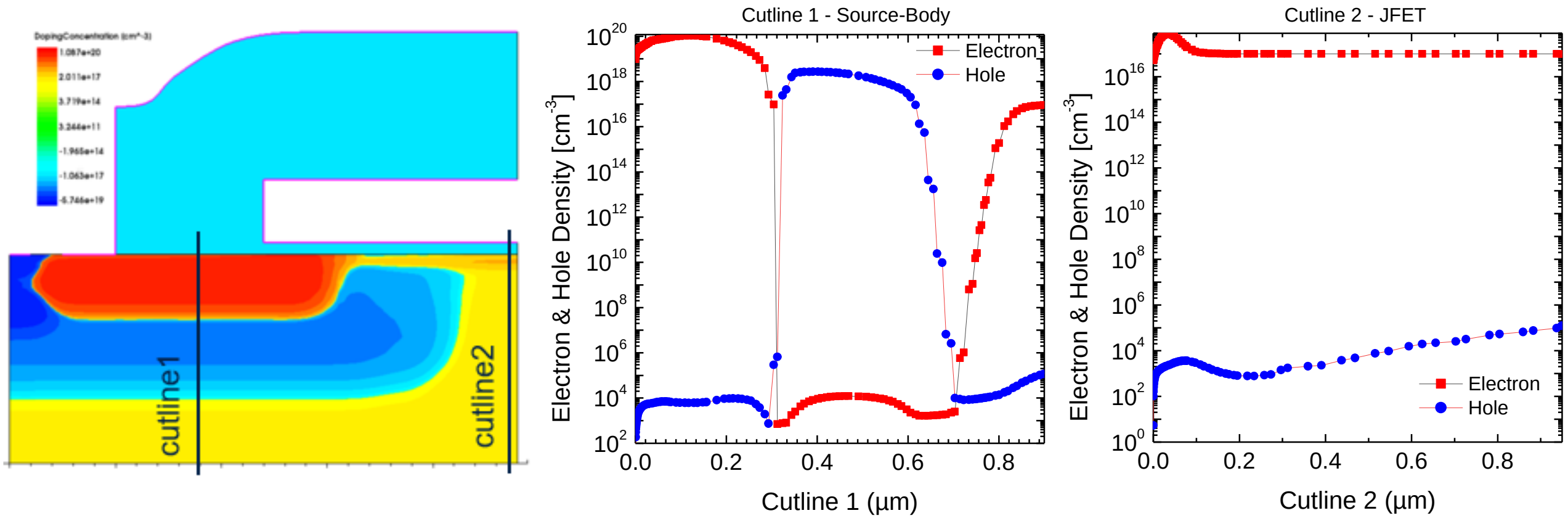

Fig. 1: TCAD free carriers simulation across the unitary cell in an ideal MOSFET.

Figs. 2a, 2b and 2c show, respectively, the AFM morphology, the SCM and the SSRM maps collected using conductive diamond coated Si tips on a MOSFET device, prepared with a bevel angle of 5° 44’, giving rise to a 10× magnification in the vertical direction. As can be noticed, the SCM image, based on local differential capacitance (dC/dV) measurements, is very sensitive to the majority carrier variations in the JFET depletion regions. On the other hand, the SSRM image, based on resistance measurements by a logarithmic current amplifier, shows clear signal variation in the gate region. To better illustrate the SSRM sensitivity and lateral resolution, the measured resistance across

the gate insulator region is depicted in Fig, 3. As can be noticed, an abrupt resistance variation across the $SiO_2$/4H-SiC interface is detected. This information can be used to determine the lateral spatial resolution (from the 10% to 90% signal variation) of about 5 nm (red box in Fig. 3). This particular information demonstrates the capabilities of the characterization technique to collect electrical information with a high spatial resolution.

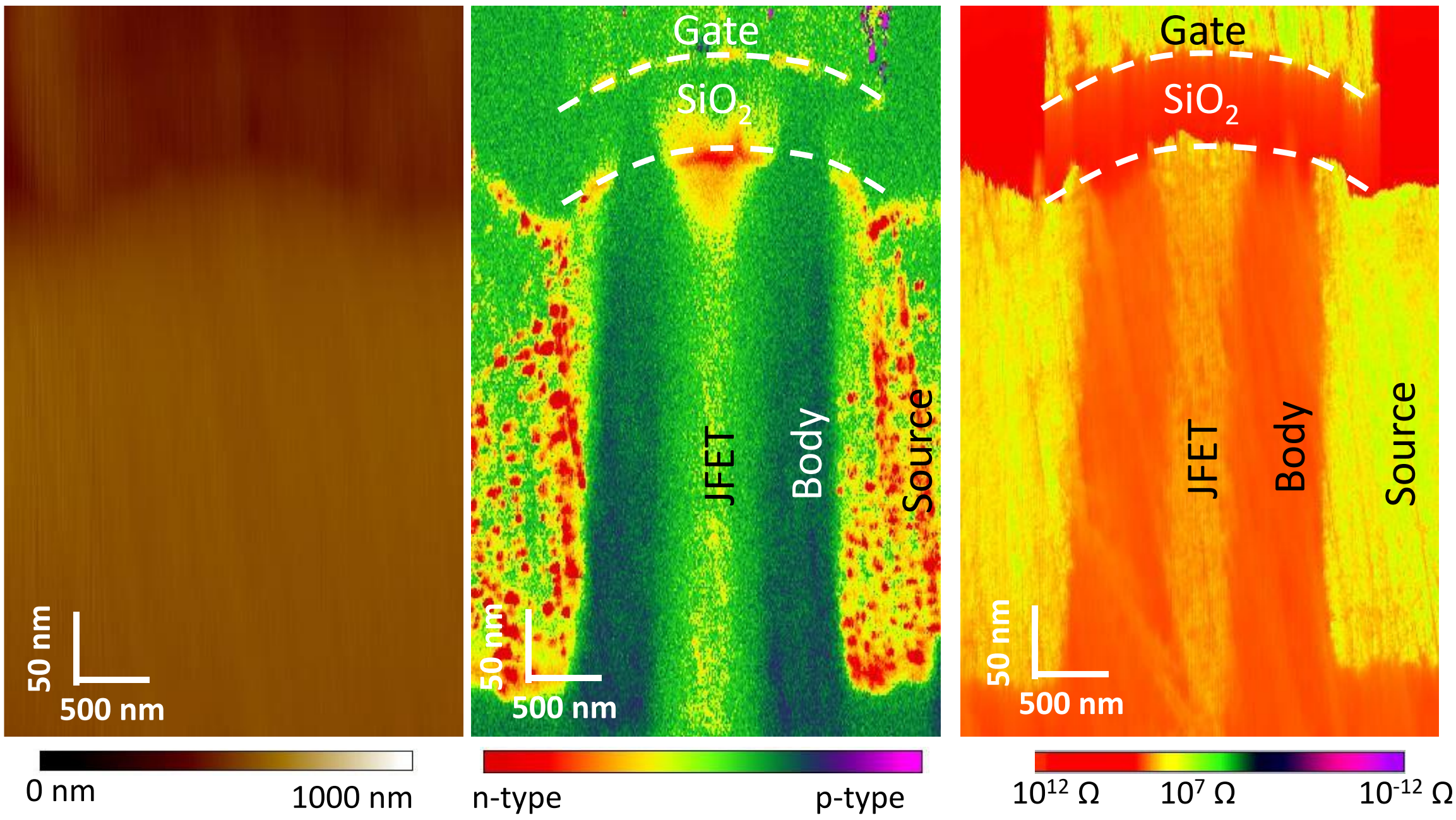

Fig. 2: (a) AFM morphology, (b) SCM and (c) SSRM signals.

A quantification of the measured resistance maps to local resistivity requires a deeper understanding of the mechanisms of current injection from the conductive tip to the differently doped 4H-SiC MOSFET regions. To this aim, local current-voltage (I-V) characteristics were collected by the SSRM logarithmic current amplifier on a single point in the Drift, Body and Source regions, as shown in Fig. 4.

The I-V collected on the drift and body regions are compatible with a forward n-type and reverse p-type Schottky-like conduction. On the other hand, the I-V collected on the source region shows a current value 4 orders of magnitude larger, compatible with current injection by thermionic field emission. Differently than for SSRM measurements on Si samples, where Ohmic contact formation between the tip and the semiconductor was achieved by applying a sufficiently high force, a non-linear behavior of I-

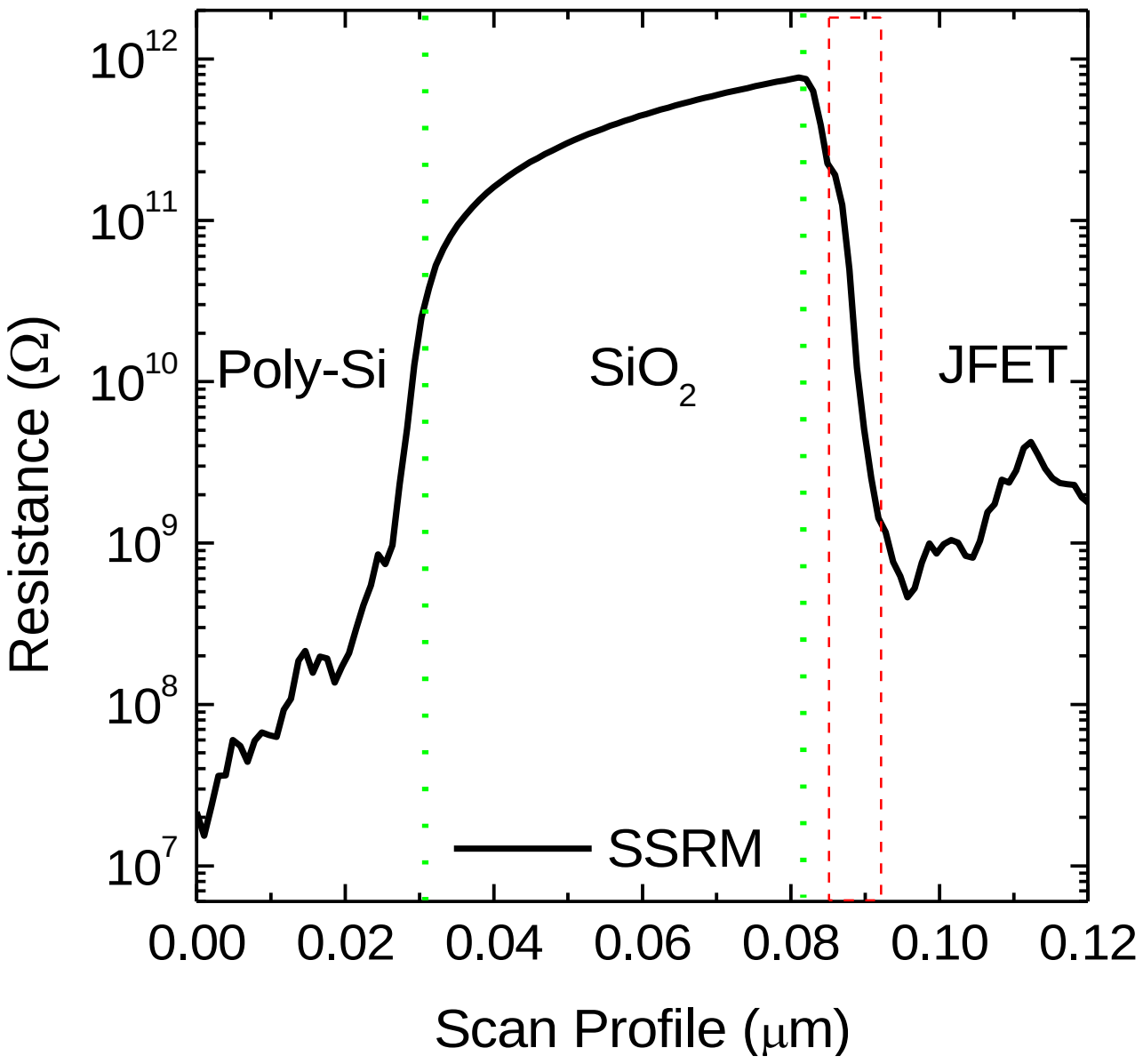

Fig. 3: SSRM profile across the gate insulator region. The green dashed lines delimitate the gate insulator. The 10% to 90% signal variation highlighted by the red dashed lines is used to demonstrate the 5 nm spatial resolution.

V curves is observed on both p- and n-type doped 4H-SiC, even on the highly doped source region. Hence, both the tip/SiC contact and spreading resistance contributions must be considered to appropriately describe current injection in 4H-SiC, in order to make a quantification of SSRM signal to resistivity map. In this respect, future work will be dedicated to SSRM investigation of 4H-SiC calibration samples with known doping types and concentrations, combined with TCAD-aided and analytical modelling of current injection at the tip/SiC junctions.

## Summary

Even though this aspect deserves further investigation, SSRM is promising for the investigation of shrunk latest MOSFETs generation in terms of lateral resolution.

An appropriate samples preparation is used to demonstrate a high spatial resolution of about 5 nm using the SSRM capabilities. Furthermore, the SCM capabilities enabled visualizing the fluctuations of charge carrier concentration across the different epitaxial and region subjected to ion implantation of the MOSFETs forming its elementary cell.

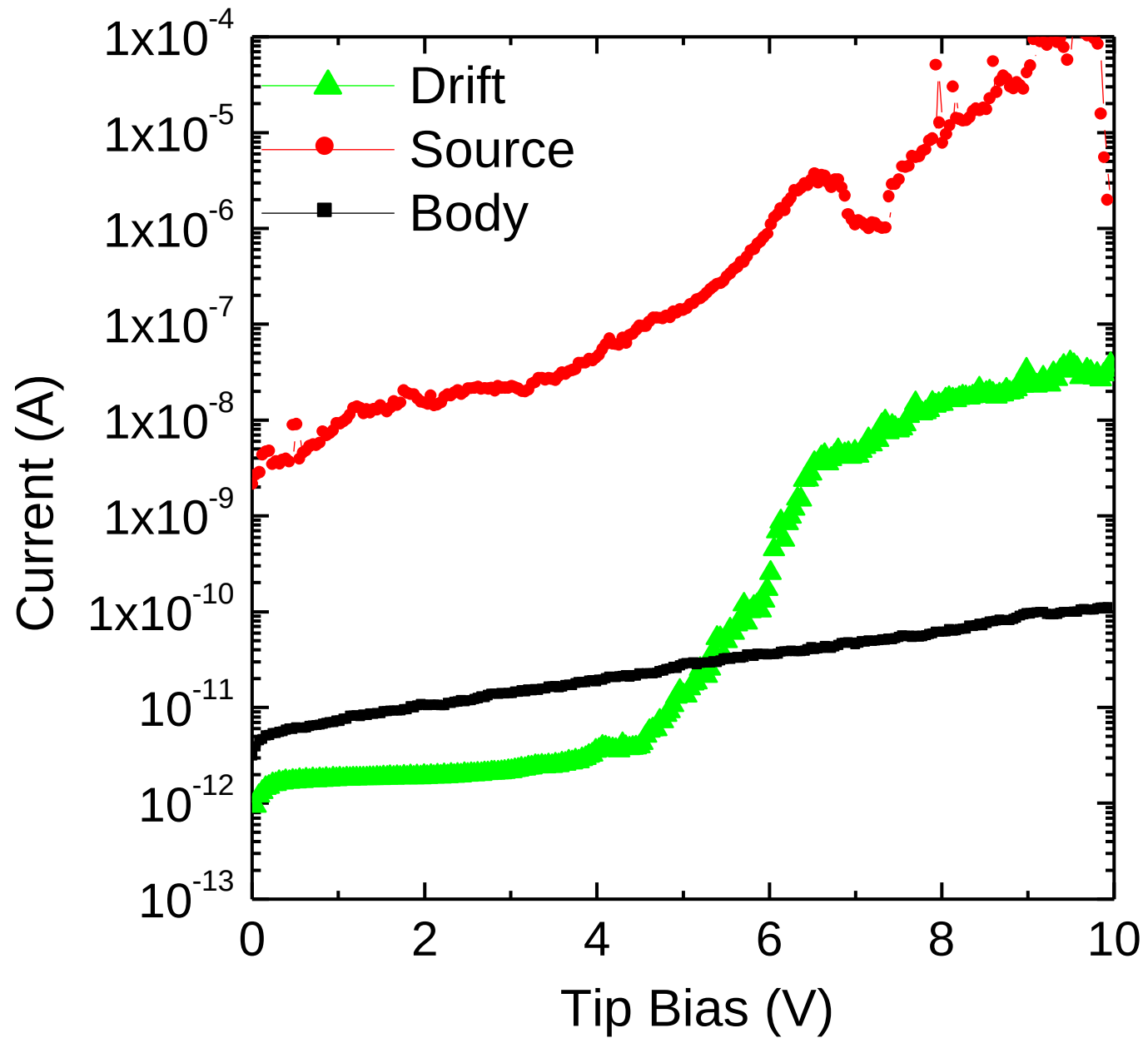

Fig. 4: Local I-V curves extracted from the SSRM signal in the Drift, Body and Source regions.

## Acknowledgements

This work is supported by the European Union Next Generation EU program through the MUR-PNRR project SAMOTHRACE (No. ECS00000022). This paper has been partially supported by AdvanSiC. The AdvanSiC project has received funding from the European Union's Horizon Europe programme under grant agreement No 101075709. Views and opinions expressed are however those of the author(s) only and do not necessarily reflect those of the European Union. Neither the European Union nor the granting authority can be held responsible for them.